# PRIVACY MANAGEMENT AND INTERFACE DESIGN FOR A SMART HOUSE

*Ana-Maria COMEAGĂ[1]*

[1]National University of Science and Technology POLITEHNICA Bucharest, Faculty of Engineering in Foreign Languages, Department of Engineering in Foreign Languages, Bucharest, Romania.

[1]ana_maria.comeaga@stud.fils.upb.ro

*Iuliana MARIN [2]*

[2]National University of Science and Technology POLITEHNICA Bucharest, Faculty of Engineering in Foreign Languages, Department of Engineering in Foreign Languages, Bucharest, Romania.

[2]ORCID ID: https://orcid.org/0000-0002-7508-1429

[2]marin.iulliana25@gmail.com

**ABSTRACT**

In today's life, more and more people tend to opt for a smart house. In this way, the idea of including technology has become popular worldwide. Despite this concept's many benefits, managing security remains an essential problem due to the shared activities. The Internet of Things system behind a smart house is based on several sensors to measure temperature, humidity, air quality, and movement. Because of being supervised every day through sensors and controlling their house only with a simple click, many people can be afraid of this new approach in terms of their privacy, and this fact can constrain them from following their habits. The security aspects should be constantly analyzed to keep the data's confidentiality and make people feel safe in their own houses. In this context, the current paper puts light on an alternative design of a platform in which the safety of homeowners is the primary purpose, and they maintain complete control over the data generated by smart devices. The current research highlights the role of security and interface design in controlling a smart house. The study underscores the importance of providing an interface that can be used easily by any person to manage data and live activities in a modern residence in an era dominated by continuously developing technology.

**INTRODUCTION**

In a century where urbanism has become a fundamental element because the population continues to double worldwide, a house is the thing without which a person can not live. This kind of urban evolution comes with many economic, social, and environmental challenges, like non-resilience, pollution, non-sustainability, and congestion [1]. Nowadays, buildings can be considered a very complex system containing various entities with interconnected structures and frameworks. If the size of the building increases, the intricacy of the involved systems also increases, and the risk of some kind of disturbance can influence people's lives and security, being a frequently met worry of a house owner. Due to this, smart buildings should diminish potential problems as much as possible and usually have some sets of solutions to control the dangers, improving the quality of life.

The main challenges of urbanization are health, education, traffic, and energy, so facing them involves the transformation of existing cities into smart cities. This concept is popular in scientific literature, describing a healthy environment that develops well-being. A smart home increases comfort, efficiency, security, and energy management by using information and communication technology, especially the Internet of Things (IoT) [2]. The idea that stays behind is based on the connection of various electronic systems and devices, which interact with people through the application interface based on algorithms.

For the implementation of smart homes, many sensors are used for multiple purposes, such as monitoring energy consumption, humidity level, air quality, and temperature. Automation based on rules and conditions can be created with the data gathered from these parameters. With the rapid growth of digitalization in the global economy, the association between the Internet of Things and humans can be efficient, creating things that some people can only dream of. For example, in a kitchen, sensors can measure the freshness of the products and send notifications when one is about to expire, maintaining the food in an optimal state [3]. Another example consists of adjusting the room's light to improve the quality of the image provided by the television and setting its color accordingly. Smart windows can have sensors that detect the brightness inside the room and set the quantity of light that can pass through them, also considering the presence of plants and the fact that they need more light to evolve.

Smart homes with specific features consist of multiple devices interconnected by an IoT system for improving life, which has become popular recently. Even so, despite that, the systems that allow the proper functionality of smart houses remain dark due to many security and privacy concerns. Most existing platforms, such as Amazon Alexa or Samsung SmartThings, usually need a third-party application that collects a certain amount of sensitive data, which can be abused in harmful ways. They rely on an optional access control model in which the user permits the platform to work with some data set. Once admitted, permissions themselves fail to control how exactly the application will use the resources, and this becomes very difficult to manage when the number of devices is increasing continuously. In this context, many techniques were tried for verifying the security and safety of the apps and improving the access control mechanisms. Currently, the platform supplier has complete control over the IoT cloud backend, meaning he can collect, store, and share users' sensor data [4].

Due to technology and innovation advancements, people are motivated to follow a comfortable lifestyle made by smart devices from the house. However, they still need to be more negligent regarding their privacy, and various cases show that a smart house can be easily hacked if it does not have a robust security system. Because of the unclear guidance of the users, the general security measures established when the system was created can be easily hacked, leaving the door open for hackers who access the entire network of the smart home automation system [5]. Smart houses are also exposed to physical attacks by interference with digital wires and installing low-quality sensors. Moreover, although the external processes seem not to be dangerous, the substantial danger is processed after firing up the gadgets. Together with these issues, the physical attacks seriously affect the integrity of data from the system because of the absence of control and evidence entertained in the security of the hardware and software systems of the devices that compose the various types of automation in the house.

The current paper describes the importance of the security of the residents for such a platform in the context of smart homes. Our approach tries to eliminate the possible attacks and vulnerabilities of the system, make it a safe place for the users, and prevent them from being victims of hackers or sharing their data without even realizing it. It also brings forward an application interface that is easy for each person and contains the necessary instructions to protect the users from unexpected dangers that can appear every time. In the next section are the research and findings utilized for a complex analysis of the proposed system, the results that improve the security and the interface design of the platform, and finally, the conclusions.

## RESEARCH AND FINDINGS

Besides the simple concept of building, it consists of all its embedded technologies and systems, which customers can modernize by choosing some facilities such as warming, illumination, entertainment, safety, conditioning systems, and ventilation. A high priority in these applications should be given to energy-saving methods, focusing on direct and indirect carbon emissions to achieve smart urbanization.

The smart home concept has advanced so much that it can provide household operations by automating the interaction between existing devices to ensure comfort, convenience, and security [6]. Although it brings many benefits to the house owner, it also comes with some privacy challenges, generally produced by smart devices, communication channels, protocols, and firmware. In recent years, many studies shed light on protecting user's data privacy.

Based on a specific IoT system, the design, development, and classification of automation are possible, and they offer the residents complete control over the energy consumption and convenience of their house. The sensors in a bright house are used for door locks, temperature, air conditioner, automation of light, kitchen stoves, water, and gas. All the data collected from the sensors are put into the central home, also called the gateway, and after that, they are sent to system control units through the Internet.

Fig. 1 presents the automation system of a usual smart house, with its primary devices that contain sensors for monitoring the parameters. These automations can be controlled from the mobile phone simply by installing the corresponding application.

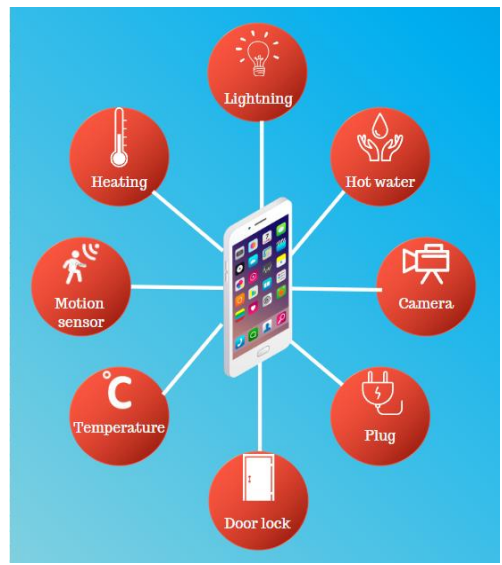

Fig. 1 Smart Home Automation System

In terms of IoT systems, there are four main categories of attacks that can happen: Physical Attacks, where the target is the hardware; Network Attacks, through which a large amount of data are extracted remotely; encryption, which means finding the encryption key and steal the data, and Software Attack, made by accessing the entire software of the platform while there is done an installing of malware, virus, phishing, injection of malicious code [7]. Apart from these, the system can be a victim of other well-known attacks. Botnets consist of a group containing remotely controlled compromised computers to execute different kinds of frauds and cyber-attacks, like stealing personal data, phishing emails, or exploiting information. They are Internet-enabled and can transfer data automatically via the Internet. These make them easily accessible from hackers and challenging to detect due to the users' lack of knowledge. Another example is a Man-in-the-Middle attack, which appears when the hacker breaks the communication between a two-end system by injecting a malicious node among the existing nodes or targeting the communication protocols inside the network. An alteration of traffic flow, a change in network topology, and the apparition of fake identities or false information can be done through this concept.

Social engineering means manipulating the user, obtaining their private data, and gaining illegal access to it. The weakness of the IoT device's passwords can easily cause brute-force attacks and steal login credentials. Ransomware is popular as one of the bad attacks in which the hackers encrypt the users' critical data and demand a pay-off to give them back. SQL injection is a security vulnerability through which someone places harmful SQL queries into the input fields of the application to operate with the database and get sensitive information [8]. This kind of attack can modify the parameters of the SQL query in such a way that they can add additional instructions or data.

A cyber attack is always carefully planned so the user does not know when he is a victim, and there are more stages to follow. The entire process is shown in Fig. 2, starting from the detection of the target, preparation for the attack, malware acquisition and deployment, to the actual attack, detection, prevail, and exploitation of the data.

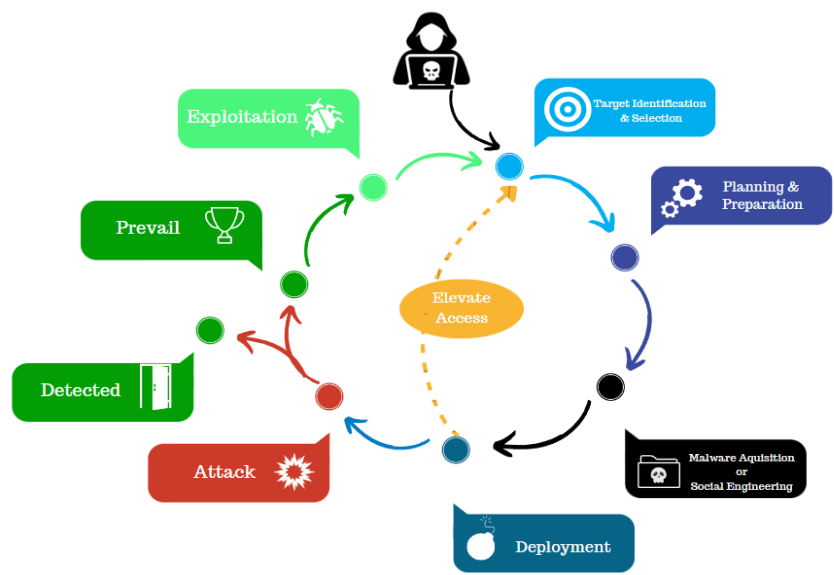

Fig. 2 Cyber Attack Cycle

The proposed system permits the design of personalized solutions to maintain security using its format, which is similar to a general graphical user interface. Home Assistant is a free software and an important home automation platform that allows the connection of different IoT protocols, such as MQTT, CoAP, and HTTP. The system's similarity with MQTT enables real-time communication of the devices and the creation of a reactive and interconnected network. It uses this protocol with its necessary security measures to reduce the exposure of the network to many forms of cyber attacks, like exploitation of the protocol behavior to undermine the devices for accessing the stored data on the server. The Constrained Application Protocol ensures the optimal communication of the devices, being a solution for in-regulated data formats and offering high security to the platform. The primary purpose of HTTP is the accessibility of the application and interaction with various web-based services. When interacting with external APIs, cloud services, or web-based applications, the HTTP's integration of the Home Assistant makes a comprehensive IoT experience possible. The system's web template engine was designed using Jinja2.

The first step is the introduction of a new location, for which a name must be chosen, together with the exact position on an interactive map or by introducing the proper values for the latitude and longitude of the place, as shown in Fig. 3. In case when the interactive map is used, the values for the latitude and longitude are automatically introduced by the program. This is useful for creating automation, knowing if someone is outside the house or moved to another zone from the system. To keep the user's privacy, the interruption of sharing the location can be done anytime, with the assumption that some functionalities of the system can be unavailable. The locations are stored in the database for a short period and are used only for the optimal operation of IoT devices.

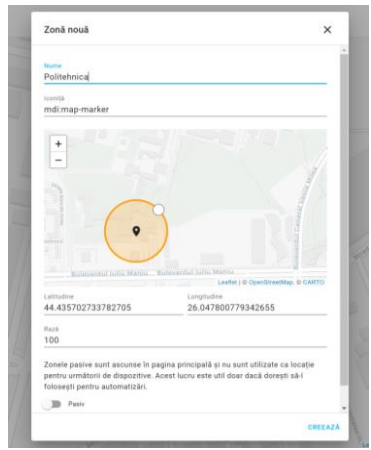

Fig. 3 Adding a new location

In the next step, after the authentication is done, the main interface will be shown to the user, and with it, the user can verify the status of the devices from the house, like indoor lightning illumination, presence of movement, door opening, and the levels of humidity, CO, smoke. From the graphical user interface, the light sensors' intensity and color can be changed, and the effects are seen immediately through the specific sensor's logo that already has the chosen color.

The interface of the mobile application for the system makes the administration of the devices accessible, and its design is illustrated in Fig. 4.

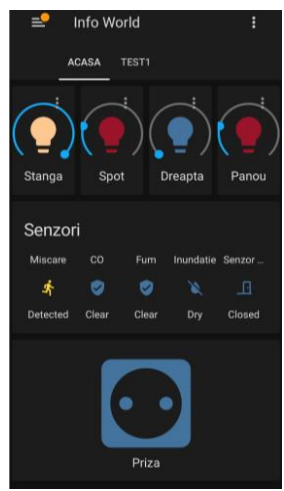

Fig. 4 Mobile application interface

The mobile application helps the user monitor all activities from the established area, even if he is anywhere else. This interface is also easy to use for the elderly living alone, being very intuitive and with the necessary instructions for managing all the devices in the house. Users can add customized panels to the graphical interface, like a panel for the person's movement or average humidity or temperature.

When the button Developer Tools is selected from the interface, followed by the Status tab, the user can see the status of an entity belonging to a device. Depending on each entity, various attributes appear on the GUI, and the value of the state can be Boolean, in the case of a motion sensor, button, or numerical, for temperature. The Set Status button is responsible for changing the status of an entity.

Users can test the behavior of events and activate automation based on triggers. This feature allows users to compare the expected result with the actual one without pressing any button and experiment with conditions and triggers. The automation process consists of three main steps. First, an event triggers it, such as activating a motion sensor or pressing a button. Common types of automation include the high presence of CO, the movement of a person or an animal, and a temperature drop. Then, the automation's conditions are examined, such as checking if the CO level is higher than in the last 3 days or if the temperature is lower than the average at a particular time. The final step involves calling a service if the conditions are met. Users can easily edit a specific button by pressing it with a pencil, and the graphical interface will immediately display the changes made, along with the type of action and the dates of an event.

## RESULTS AND DISCUSSIONS

As you navigate to the application menu, you'll find the reports, which you can generate by pressing the Log button. These detailed reports show the event's type, such as panel off or motion detection, and the associated entity. You can select a specific period to view the events and decide how long to store them in the system. These reports can be downloaded onto your devices, allowing you to analyze them conveniently. This control over the reports empowers you to observe how the entities modify over a month or year and manage perturbing factors easily, enhancing your smart house's quality of life.

The system's security is one of the most crucial factors that influence the platform's development over time. The entire algorithm is at risk in a cyber attack, potentially leading to unexpected behavior. As shown in Table 1, for certain types of attacks and scenarios, it's essential to have a set of prevention measures and to train you, the user, with specific instructions. Understanding when you might be a victim of such an incursion is key. Your role in preventing these attacks is significant, as it directly impacts improving the security system tailored to each scenario and similar ones. Therefore, you need to be vigilant with the permissions you grant to the application, the complexity of passwords you set, the authentication system, and the integrity of the sensors.

Table 1. Prevention measures for the most common cyber attacks

| Attack type | Scenario | Prevention |
|---|---|---|
| Physical attack | The dook locker is forced | Facial recognition of the residents |
| Social Engineering | The level of CO or smoke is increased | Alerts for users and system's manager |
| DDoS attack | The house heating system is affected | Web application firewall |
| Ransomware | The motion sensors are stopped | Strong password to start the system |
| Man-in-the-Middle attack | The connection between the devices is defective | Multiple factor authentication |

In the event of a physical attack, such as when the door lock is tampered with from the outside, and the sensors could be damaged, one effective preventative measure is installing facial recognition for the residents. This feature ensures that if an unrecognized individual attempts to enter the house, an alert is immediately sent to the users, thereby enhancing the security of the residents and their property. In the case of a social engineering attack, where the parameters of CO or smoke are manipulated from the outside with the user's unwitting consent through an unauthorized update, the system responds by sending a notification to the user and a designated contact. This notification informs them that the parameters have exceeded the safe threshold, and a message is also sent to the system's manager, who is responsible for investigating the source of the attack and restoring the platform to a safe state.

There is a risk of hackers infiltrating the system and disrupting the heating process throughout the house. They could initiate a DDoS attack, forcing the application to restart repeatedly and thus preventing the heating system from starting. To counteract this, a web application firewall can be employed. This tool filters analyzes, and blocks malicious HTTP traffic between the application and the Internet, ensuring the uninterrupted functioning of the home's heating system. In the event of a Ransomware attack, where the motion sensors could be turned off with no possibility of restarting, the system responds by requiring the homeowner to manually activate the sensor by entering a password into the application interface, which must be changed monthly. Suppose the system falls victim to a Man-in-the-Middle attack, where all devices are affected, and the connection between them is disrupted. In that case, the recommended solution is to configure the platform for multi-factor authentication and regularly verify the network status to which they are connected.

A key component in the optimal functionality of the application is the user interface. The design must be straightforward and understandable to all users, featuring clear instructions and specific images. This is vital for users to understand each button's purpose and control the devices in their homes without needing assistance. Table 2 presents some of the most common issues users have encountered and how the platform has been updated to make their experience more effortless.

Table 2. Development of the system based on users' issues

| Encountered problem | Solutions |
| --- | --- |
| Memorizing difficulties | - similar functions for different tasks<br>- descriptive text under each button |
| Cognitive limitations | - visual tutorials for users<br>- navigation paths easy to understand |
| Visual impairments | - zoom the page as much as needed<br>- audio element for each functionality |
| Color blinds | - different color used<br>- color distinguishable for color blind people |

People who had problems using the platform were memorizing the steps that needed to be followed for each task, some cognitive limitations, visual impairments, and color blinds. Because of this, the platform was improved with descriptive text for every button, visual tutorials, audio elements, and various colors that can be visualized easily. Now, the interface is more interactive, the users clearly understand what it does, and they have permanent support for each issue. Feedback can be sent whenever they want, so the application is continuously updated for people's needs. The interface offers the house owner complete control over the devices in the house, and the automation can be manually activated from it. The simplicity helps a lot of older adults, especially those who live alone, without the help of relatives.

## CONCLUSIONS

The proposed system underscores the importance of personalized security solutions within smart homes, leveraging a user-friendly interface reminiscent of a general GUI. This interface facilitates seamless integration with IoT protocols, including MQTT, CoAP, and HTTP, ensuring efficient communication and interoperability across diverse devices and platforms. By adopting these protocols, the system enables real-time device communication and network security measures, vital for maintaining the integrity and confidentiality of user data.

The user interface offers comprehensive features, allowing users to set up new locations, authenticate, and access device status information effortlessly. Users can monitor various parameters such as indoor lighting, movement, door status, and environmental conditions like humidity, CO, and smoke levels through the graphical interface. Additionally, users can adjust the intensity and color of light sensors directly from the interface, with changes reflected instantly.

Mobile accessibility via the system's mobile application further enhances user convenience, enabling remote monitoring and control of smart home devices from anywhere. This mobile interface is designed to be intuitive and user-friendly, catering to the needs of elderly users and caregivers alike. Customizable panels within the graphical interface allow users to personalize their smart home experience, tailoring it to their preferences and requirements.

In terms of security, the system implements robust measures to prevent common cyber-attacks, ensuring the integrity and confidentiality of user data. By proactively addressing potential security threats, the system aims to provide users with a safe and reliable smart home environment.

The proposed system offers a comprehensive and user-friendly smart home solution, combining advanced IoT functionalities with robust security measures and intuitive interface design. By prioritizing user experience and security, the system aims to deliver a seamless and personalized smart home experience for users of all backgrounds and abilities.

**REFERENCES**

1. Alghamdi, M. (2024). Smart city urban planning using an evolutionary deep learning model. Soft Computing, 28(1), 447-459.
2. Sinaga, D. C. P., Tampubolon, G. J., & Ndruru, I. (2024). IMPLEMENTATION OF A SMART HOME BASED ON INTERNET OF THINGS USING CISCO PACKET TRACER. Journal of Computer Networks, Architecture and High Performance Computing, 6(1), 407-418.
3. Mukhopadhyay, S., & Suryadevara, N. K. (2023). Smart Cities and Homes: Current Status and Future Possibilities. Journal of Sensor and Actuator Networks, 12(2), 25.
4. Zavalyshyn, I., Santos, N., Sadre, R., & Legay, A. (2020, December). My house, my rules: A private-by-design smart home platform. In MobiQuitous 2020-17th EAI International Conference on Mobile and Ubiquitous Systems: Computing, Networking and Services, 273-282.
5. Aldahmani, A., Ouni, B., Lestable, T., & Debbah, M. (2023). Cyber-security of embedded IoTs in smart homes: challenges, requirements, countermeasures, and trends. IEEE Open Journal of Vehicular Technology, 4, 281-292.
6. Alalade, E. D., Mahyoub, M., & Matrawy, A. (2024). Privacy Engineering in Smart Home (SH) Systems: A Comprehensive Privacy Threat Analysis and Risk Management Approach. arXiv preprint arXiv:2401.09519.
7. Sivasankari, N., & Kamalakkannan, S. (2022). Detection and prevention of man-in-the-middle attack in iot network using regression modeling. Advances in Engineering Software, 169, 103126.
8. Noman, H. A., & Abu-Sharkh, O. M. (2023). Code Injection Attacks in Wireless-Based Internet of Things (IoT): A Comprehensive Review and Practical Implementations. Sensors, 23(13), 6067.